\documentclass[12pt]{article}

\usepackage{amsmath,amssymb,amsthm,graphicx}

\newcommand{\n}{\noindent}

\newcommand{\ed}{\end{document}}
\newcommand{\be}{\begin{equation}}
\newcommand{\ee}{\end{equation}}

\begin{document}
\begin{center}
\large{\textbf{\textbf{Noncommutative Spacetime in Very Special Relativity}}}\\
\end{center}
\begin{center}
Sudipta Das$^{a,}$\footnote{E-mail: sudipta.das\_r@isical.ac.in},
Subir Ghosh$^{a,}$\footnote{E-mail: sghosh@isical.ac.in} and
Salvatore Mignemi$^{b, }$\footnote{E-mail: smignemi@unica.it}\\
$^a$Physics and Applied Mathematics Unit, Indian Statistical
Institute\\
203 B. T. Road, Kolkata 700108, India \\
$^b$Dipartimento di Matematica, Universita' di Cagliari\\
     viale Merello 92, 09123 Cagliari, Italy
         and INFN, Sezione di Cagliari\\
\end{center}\vspace{1cm}

\begin{center}
{\textbf{Abstract}}
\end{center}

\n  Very Special Relativity (VSR) framework, proposed by Cohen and
Glashow \cite{cohen}, demonstrated that a proper subgroup of the
Poincar\'e group, (in particular $ISIM(2)$), is sufficient to
describe the spacetime symmetries of the so far observed physical
phenomena. Subsequently a deformation of the latter, $DISIM_b(2)$,
was suggested by Gibbons, Gomis and Pope \cite{gibbons}.
In the present work, we introduce a novel
Non-Commutative (NC) spacetime structure, underlying the
$DISIM_b(2)$. This allows us to construct explicitly the
$DISIM_b(2)$ generators, consisting of a sector of Lorentz rotation generators
and the translation generators. Exploiting the Darboux map
technique, we construct a point particle Lagrangian that lives in
the NC phase space proposed by us and satisfies the modified
dispersion relation proposed by Gibbons et. al. \cite{gibbons}.
It is interesting to note that in our formulation the momentum
algebra becomes non-commutative.

\vspace{1cm}

\n Fax (Subir Ghosh): +91(33) 25773026\\
Telephone (Subir Ghosh): +91(33) 25753362

\newpage

\section{Introduction}

Einstein's Special Relativity (SR) theory invokes that all the
physical theories as well as all the observables remain invariant
or covariant under the Poincar\'e symmetry. Poincar\'e symmetry is
implemented by the Poincar\'e group, consisting of the Lorentz
transformations (boosts plus rotations) along with the spacetime
translations. In mathematical terminology, the Poincar\'e group is
the isometry group of the $(3+1)$-dimensional Minkowski spacetime.
In the particle physics sector, at low energy scales (QED + QCD),
parity (P), charge conjugation (C) and time reversal (T) are
individually good symmetries of nature. However, for higher
energies, there is evidence of CP violation. From a purely
theoretical point of view, one may even consider the breaking of
Poincar\'e symmetry at such high energy scales. So it might be
possible to describe the spacetime symmetry of all the observed
physical phenomena considering some proper subgroups of the
Lorentz group along with the spacetime translations. The
underlying criterion is that these subgroups, together with either
of P, T or CP, can be enlarged to the full Lorentz group. The
generic models based on these smaller subgroups are restricted by
the principle of Very Special Relativity (VSR), proposed by Cohen
and Glashow \cite{cohen}. The authors of \cite{cohen} identified
these VSR subgroups up to isomorphism as $T(2)$ ($2$-dimensional
translations), $E(2)$ ($3$-parameter Euclidean motion), $HOM(2)$
($3$-parameter orientation preserving transformations) and
$SIM(2)$ ($4$-parameter similitude group). The semi-direct product
of the $SIM(2)$ group with the spacetime translation group gives a
$8$-dimensional subgroup of the Poincar\'e group called $ISIM(2)$.

Stringent observational bounds on the CP violation put a
constraint that the deviation of VSR from SR should be very small.
Thus it is very difficult to observe the effects of VSR theory in
the physical scenarios. Fortunately, in case of $SIM(2)$, there are
no invariant vector or tensor fields (the so called ``spurion
fields"), although an invariant null direction is present.
This particular VSR theory therefore appears to be
compatible with all the current experimental limits on violations
of Poincar\'e invariance \cite{cohen, dunn}.

VSR phenomenology has been investigated in \cite{das}. In attempting
to construct a quantum field theory based on the above VSR subgroups,
Sheikh-Jabbari and Tureanu \cite{mm} noticed a problem: all the
above proper subgroups allow only one-dimensional representations
and hence can not represent the nature faithfully. However, the
authors of \cite{mm} provided an ingenious resolution of the
representation problem: they generalize the normal products of
operators as deformed or twisted coproducts \cite{ch}. The resulting novel forms
of NC spacetimes were further studied as arena of generalized particle dynamics \cite{sgp}.

In their work, Gibbons et al \cite{gibbons} proposed that gravity
may be incorporated in the $ISIM(2)$ invariant VSR theory by
taking deformations of the $ISIM(2)$ group such that the
spacetime translations become non-commutative. In fact, in
\cite{gibbons}, the authors showed that there exists a one-parameter
family of continuous deformations of the group $ISIM(2)$, which
they denoted by $DISIM_b(2)$. For any values of the deformation
parameter $b$, the group $DISIM_b(2)$ is an $8$-dimensional
subgroup of the $11$-dimensional Weyl group (semi-direct product
of the Poincar\'e group with the dilation). Interestingly, Gibbons
et. al. \cite{gibbons} arrived at a $DISIM_b(2)$-invariant point
particle Lagrangian $L$ for a particle with mass $m$:
\be L = -m (-\dot{x}^2)^{\frac{1-b}{2}} (-\alpha \dot{x})^{b},
\label{gibblag} \ee
which is of the Finsler form, and was first proposed by Bogoslovsky
in a different context \cite{bogos}. This observation prompted the
authors of \cite{gibbons} to argue that deforming VSR theory one
arrives at Finsler geometry, thus maintaining the commutativity of
spacetime translations. The constant vector $\alpha_\mu$ in
(\ref{gibblag}) denotes the invariant direction mentioned before
that breaks the full Poincar\'e invariance. The induced dispersion
relation obtained from (\ref{gibblag}) is
\be p^2 + m^2(1-b^2) \left(\frac{(-\alpha
p)}{m(1-b)}\right)^{\frac{2 b}{1+b}}=0. \label{g2} \ee

In this perspective our work bears a special significance, since we
have explicitly shown that starting from a first-order Lagrangian
with an underlying noncommutative spacetime structure, one can
still arrive at the same Finsler Lagrangian (\ref{gibblag}). {\it{In
particular, the momentum algebra in our model are indeed
noncommutative.}}

Furthermore, Finsler geometry as a natural generalization of Riemann geometry may
provide new sight on modern physics. The models based on Finsler
geometry are claimed to explain recent astronomical observations which
are not explained in the framework of Einstein's gravity. For example,
the flat rotation curves of spiral galaxies can be deduced naturally
without invoking dark matter \cite{f1} and the anomalous acceleration
in solar system observed by Pioneer 10 and 11 spacecrafts can be accounted for
\cite{f2}.

Our paper is organized as follows: In Section 2, we describe the
$DISIM_b(2)$ algebra given by Gibbons \cite{gibbons}. In Section
3, the underlying non-commutative spacetime structure representing
the $DISIM_b(2)$ algebra is derived explicitly. This is one of the
major result of our paper. In Section 4, we introduce a
reparametrization invariant first-order (phase space) point-particle Lagrangian
and it is shown in detail that starting with this Lagrangian,
the non-commutative spacetime structure specified above
can be exactly reproduced using the well-known method of
Dirac constraint analysis \cite{dirac}. In Section 5,
we proceed to formulate a second-order (coordinate space) particle
Lagrangian. Notably we recover the second-order Lagrangian
(\ref{gibblag}), as proposed in \cite{bogos, gibbons}. Our paper ends
with a conclusion in Section 6.

\section{Deformed $ISIM(2)$ or $DISIM_b(2)$ Algebra}

In \cite{gibbons}, the authors have considered the VSR theory
based on $ISIM(2)$, one of the subgroups suggested by Cohen and
Glashow \cite{cohen}. In their attempt to incorporate gravity into this VSR
framework, the authors of \cite{gibbons} further considered a
$1$-parameter family of deformations ($b$ being the deformation
parameter), called $DISIM_b(2)$, under which the VSR theory is
invariant. The group $DISIM_b(2)$ has the following $eight$
generators: \be J_{+ -}, J_{+ i}, J_{i j}, p_{+}, p_{-},
p_{i}~~;~~(i, j) = 1, 2 \label{gen} \ee where $J$-s are the
rotation and boost generators and $p$-s are the generators for the
translations. Here we used the usual light-cone coordinates $$
x_{\pm} = \frac{1}{\sqrt{2}} \left(x_0 \pm x_3 \right)~,~x_i,~~
i=1, 2 $$ and the metric $g_{\mu \nu}$ is of the form $$ g_{\mu
\nu} = diag(1, -1, -1, -1). $$

The $DISIM_b(2)$ algebra is explicitly given by \cite{gibbons}:
\be [J_{+ -},p_{\pm}] = -(b \pm 1) p_{\pm}~~,~~[J_{+ -},p_i] = -b
p_i~~,~~[J_{+ -},J_{+ i}] = -J_{+ i}~, $$$$ [J_{1 2},p_i] =
\epsilon_{i j} p_j~~,~~[J_{1 2},J_{+ i}] = \epsilon_{i j} J_{+
j}~~,~~[J_{+ i},p_-] = p_i~~,~~[J_{+ i},p_j] = g_{i j} p_+,
\label{algibbons} \ee
where we have chosen $\epsilon_{12}=-1$.

In (\ref{algibbons}), if the deformation parameter $b$ vanishes,
we recover the usual canonical algebra. In \cite{gibbons}, on the
basis of this non-canonical $DISIM_b(2)$ algebra
(\ref{algibbons}), the Finsler Lagrangian (\ref{gibblag}) was
obtained. However, the symplectic structure between the phase
space variables $(x, p)$ was not explicitly specified in
(\ref{algibbons}).

Our main motivation stems from the fact that the non-canonical
algebra of the generators postulated in \cite{gibbons} appears in
a purely algebraic way. Hence our aim is to understand the
appearance of this algebra at a more physical and fundamental
level. In fact, from our previous experience, we know that a
generalized version of Nambu-Goto type action (as in
(\ref{gibblag})) and modified dispersion relation (as in
(\ref{g2})) are usually connected to an NC phase space (example in
other contexts can be found in \cite{darb}).

In the following sections we show that the algebra
(\ref{algibbons}) can be induced by a new and interesting form of
Non-Commutative (NC) spacetime (or more accurately NC phase
space). This NC phase space algebra allows us to explicitly
construct the $DISIM_b(2)$ generators in terms of phase space
degrees of freedom. Finally, armed with this novel phase space, we
go on to construct a point particle Lagrangian in a systematic and
physically transparent way. In the process we recover
the Lagrangian proposed by Gibbons et. al. in \cite{gibbons}.
An added bonus in our scheme, which makes the
study of $DISIM_b(2)$ \cite{gibbons} all the more attractive is
that the momenta $p_{\mu}$ become noncommutative.

\section{The Underlying NC Phase Space Algebra}

In this section we will disclose the novel NC phase space algebra
which will turn out to be fairly involved, but we will
see that it can be recast in a more convenient and symmetric form.
Since there is no unique way to recover the algebra, we fall
back to the canonical prescription for the angular momentum structure
and postulate the generators to be
\be J_{\mu \nu} = x_{\mu} p_{\nu}
- x_{\nu} p_{\mu} \label{j}. \ee
Obviously, to satisfy the deformed algebra (\ref{algibbons}) a
non-canonical or NC $(x, p)$ phase space algebra is required. We
posit it to be of the following form:
\be [x_+,p_-] = 1-\frac{b}{2}~~,~~[x_-,p_+] = 1+\frac{b}{2}~~,~~[x_-,p_-] =
\frac{b}{2} \left(-\frac{x_-}{x_+} + \frac{p_-}{p_+} \right), $$$$
[x_+,p_+] = 0~~,~~[x_-,x_+] = -\frac{b}{2}
\frac{x_+}{p_+}~~,~~[p_-,p_+] = -\frac{b}{2}
\frac{p_+}{x_+}~~,~~[x_-,p_i] = \frac{b}{2} \frac{p_i}{p_+}, $$$$
[p_-,x_i] = \frac{b}{2} \frac{x_i}{x_+}~~,~~[x_-,x_i] =
-\frac{b}{2} \frac{x_i}{p_+}~~,~~[p_-,p_i] = -\frac{b}{2}
\frac{p_i}{x_+}, $$$$ [x_+,p_i] = [p_+,x_i] = [x_+,x_i] =
[p_+,p_i] = 0~~,~~[x_i,p_j] = g_{ij}. \label{comm} \ee
The above algebra satisfies the Jacobi identity. It is straightforward
to convince oneself that the algebra (\ref{comm}) together with
the definition of angular momentum (\ref{j}) reproduces the
$DISIM_b(2)$ algebra (\ref{algibbons}). This is the first
important result of our paper.

Our next objective is to search for a canonical representation of
the NC algebra (\ref{comm}) which will be vital in our systematic
derivation of the particle Lagrangian. That this map (between NC
and canonical variables) is in principle derivable follows
from Darboux's theorem, which states that it is possible
at least locally to construct an invertible map between NC and
canonical (commutative) phase space variables. It should also be
stressed that in general the Darboux map might be very involved
and difficult to construct explicitly, and we are not aware of a
unique or systematic method for deriving this map.

In the present case, the explicit form of the
Darboux map between the noncommutative
phase space variables $(x_{\mu}, p_{\mu})$ and the canonical phase
space variables $(X_{\mu},P_{\mu})$ with
$$[X_\mu,P_\nu]=g_{\mu\nu }~~~,~~~[X_\mu,X_\nu] = [P_\mu,P_\nu]=0$$
is given by
\be x_+ = X_+~~,~~x_i = X_i~~,~~p_+ = P_+~~,~~p_i = P_i, $$$$ x_-
= X_- + \frac{b}{2} \frac{(XP)}{P_+}~~~,~~~p_- = P_- - \frac{b}{2}
\frac{(XP)}{X_+} \label{darb} \ee where $(X P) = X_{\mu}P^{\mu}$.
The inverse Darboux map can be readily obtained,
\be X_+ = x_+~~,~~X_i = x_i~~,~~P_+ = p_+~~,~~P_i = p_i,
$$$$ X_- = x_- - \frac{b}{2} \frac{(x p)}{p_+}~~~,~~~P_- = p_- +
\frac{b}{2} \frac{(x p)}{x_+}. \label{invdarb} \ee
For later convenience, we stick to a manifestly covariant framework
by introducing two constant null vectors
$\alpha_{\mu}=(\frac{1}{\sqrt{2}}, 0, 0, -\frac{1}{\sqrt{2}})$ and
$\beta_{\mu} = (\frac{1}{\sqrt{2}}, 0, 0, \frac{1}{\sqrt{2}})$,
satisfying the relations $\alpha^2 = \beta^2 = 0~,~(\alpha \beta)
= 1$. Thus in our notation, $ (\alpha x) = \frac{x_0 +
x_3}{\sqrt{2}} \equiv x_{+},~~(\beta x) = \frac{x_0 -
x_3}{\sqrt{2}} \equiv x_{-} $. With the help of the above, the
Darboux and inverse Darboux maps (\ref{darb}, \ref{invdarb}) can
be written in the covariant form:
\be x_{\mu} = X_{\mu} + \frac{b}{2} \frac{(X P)}{(\alpha P)}
\alpha_{\mu}~~~,~~~p_{\mu} = P_{\mu} - \frac{b}{2} \frac{(X
P)}{(\alpha X)} \alpha _{\mu} \label{covdarb} \ee
\be X_{\mu} = x_{\mu} - \frac{b}{2} \frac{(x p)}{(\alpha p)}
\alpha_{\mu}~~~,~~~P_{\mu} = p_{\mu} + \frac{b}{2} \frac{(x
p)}{(\alpha x)} \alpha_{\mu}. \label{covinvdarb} \ee
Furthermore, we rewrite our new NC phase space algebra
(\ref{comm}) in a covariant form as well,
\be [x_{\mu},x_{\nu}] = \frac{b}{2 (\alpha p)} (\alpha_{\nu}
x_{\mu} - \alpha_{\mu} x_{\nu})~~~,~~~[p_{\mu},p_{\nu}] =
\frac{b}{2 (\alpha x)}(\alpha_{\nu} p_{\mu} - \alpha_{\mu}
p_{\nu}) $$$$ [x_{\mu},p_{\nu}] = g_{\mu \nu} - \frac{b}{2 (\alpha
x)} \alpha_{\nu} x_{\mu} + \frac{b}{2 (\alpha p)} \alpha_{\mu}
p_{\nu}. \label{covcomm} \ee
The expressions (\ref{covcomm}) give the underlying
non-commutative phase space structure of the
$DISIM_b(2)$-invariant Very Special Relativity
(VSR) theory. We observe a very pleasing
symmetrical structure in (\ref{covdarb},
\ref{covinvdarb}, \ref{covcomm}) under the exchange
between $x_\mu $ and $p_\mu $.

The infinitesimal transformations of $x_\mu $ and $p_\mu $ and the
NC Lorentz algebra follow easily,
\be [J_{\mu \nu}, x_{\rho}] = g_{\mu \rho} x_{\nu} - g_{\nu \rho}
x_{\mu} - \frac{b}{2 (\alpha p)} x_{\rho} \left(\alpha_{\mu}
p_{\nu} - \alpha_{\nu} p_{\mu} \right) - \frac{b}{2 (\alpha x)}
\left(\alpha_{\mu} x_{\nu} - \alpha_{\nu} x_{\mu} \right) x_{\rho}
$$$$ [J_{\mu \nu}, p_{\rho}] = g_{\mu \rho} p_{\nu} - g_{\nu \rho}
p_{\mu} + \frac{b}{2 (\alpha p)} p_{\rho} \left(\alpha_{\mu}
p_{\nu} - \alpha_{\nu} p_{\mu} \right) + \frac{b}{2 (\alpha x)}
\left(\alpha_{\mu} x_{\nu} - \alpha_{\nu} x_{\mu} \right) p_{\rho} \label{alnon} \ee
\be [J_{\mu \nu}, J_{\rho \sigma}] = g_{\mu \rho} J_{\nu \sigma} -
g_{\nu \rho} J_{\mu \sigma} + g_{\mu \sigma} J_{\rho \nu} - g_{\nu
\sigma} J_{\rho \mu} $$$$ - \frac{b^2}{4 (\alpha x) (\alpha p)}
\left(\alpha_{\mu} \alpha_{\rho} J_{\nu \sigma} - \alpha_{\nu}
\alpha_{\rho} J_{\mu \sigma} + \alpha_{\mu} \alpha_{\sigma}
J_{\rho \nu} - \alpha_{\nu} \alpha_{\sigma} J_{\rho \mu}\right).
\label{al} \ee It is interesting to note that the Lorentz algebra
(\ref{al}) is deformed at $O(b^2)$ only and all the above
relations will reduce to (4) in the relevant sector of the
algebra, to match \cite{gibbons}.

\section{Particle Lagrangian: First-Order Form}

Enough of kinematics! Let us now consider dynamics in this NC
phase space scenario. The first and foremost problem is to
construct a point particle model that will live in this new phase
space described by (\ref{covcomm}). Since the dynamics is in phase
space, a first-order form is most suitable for our purpose. Once
the Darboux map relating the NC phase space variables to the
corresponding canonical variables is derived, one can conveniently
construct a dynamical model in the NC phase space
applying the following prescription: start with a known
(canonical) action, exploit the Darboux map to express the action
in terms of the NC phase space variables and then
study the dynamics. Though in principle the Darboux map exists, in
practice it is not always possible to derive such a Darboux map.
In our work, we are able to find out the explicit expression for
the Darboux map which helps us to derive the Lagrangian for this
VSR model. This scheme has been exploited before in \cite{darb}.

We want to define a Lagrangian $L$ such that we can reproduce the
above non-commutative structure (\ref{covcomm}) from its kinetic
part. So we start with
\be L = P_{\mu} \dot {X}^{\mu} -
\frac{\lambda}{2} \left( P^2 + A (-\alpha P)^{\frac{2
b}{1+b}}\right), \label{flagg} \ee
where $X_{\mu}$ and $P_{\mu}$ represent canonical phase space
variables, $ \dot{X}_{\mu} = \frac{d X_{\mu}}{d \tau}$ and $A$ is
an arbitrary constant. The Lagrange multiplier $\lambda $ enforces
the mass-shell constraint. The arbitrary constant $A$ is
determined from the mass-shell constraint as $ A = -
\frac{P^2}{(-\alpha P)^{\frac{2 b}{1+b}}} $, since the ratio $
\frac{P^2}{(-\alpha P)^{\frac{2 b}{1+b}}} $ is invariant under the
$DISIM_b(2)$ algebra given by (\ref{algibbons}) and (\ref{alnon}),
\be \left[J_{\mu \nu}, \frac{P^2}{(-\alpha P)^{\frac{2
b}{1+b}}}\right] = 0~~;~~\mu, \nu = +, -, i. \label{inv} \ee
It is important to keep in mind that $J_{\mu\nu}=x_\mu p_\nu
-x_\nu p_\mu$, as defined before, and one way of computing
(\ref{inv}) is to convert $J_{\mu \nu}$ to the canonical
coordinates by using the Darboux map (\ref{covdarb}). Obviously
$J_{\mu \nu}$ has a non-canonical structure in $X-P$ (canonical)
space: $$ J_{\mu \nu} = X_{\mu} P_{\nu} - X_{\nu} P_{\mu} +
\frac{b (X P)}{2( \alpha P)} \left(\alpha_{\mu} P_{\nu} -
\alpha_{\nu} P_{\mu}\right) + \frac{b (X P)}{2( \alpha X)}
\left(\alpha_{\mu} X_{\nu} - \alpha_{\nu} X_{\mu}\right). $$ The
Lagrangian (\ref{flagg}) is exactly of the same form as given in
\cite{gibbons}.

Applying the inverse Darboux map (\ref{covinvdarb}) on
(\ref{flagg}) we get the required first-order Lagrangian:
\be L = p_{\mu} \dot{x}^{\mu} + \frac{b (x p) (\alpha \dot{x})}{2 (\alpha
x)} + \frac{b (x p) (\alpha \dot{p})}{2 (\alpha p)} -
\frac{\lambda}{2} \left( p^2 + \frac{b (x p) (\alpha p)}{(\alpha
x)} + A (-\alpha p)^{\frac{2 b}{1 + b}}\right). \label{flag} \ee
Before proceeding further, let us quickly recall the main features
of Dirac's constraint analysis in Hamiltonian formulation. One
starts by computing the conjugate momentum $p=\frac{\partial
L}{\partial \dot q}$ of a generic variable $q$ and identifies the
relations that do not contain time derivatives as (Hamiltonian)
constraints. A constraint is classified as First Class when it
commutes with all other constraints and a set of constraints are
Second Class when they do not commute. First Class constraints
generate gauge invariance which is not of our concern here.
However, for systems containing second class constraints, one has
to replace the Poisson brackets by Dirac brackets, to properly
incorporate the Second Class constraints. If $([\psi_{\rho}^{i},
\psi_{\sigma}^{j}]^{-1})$ is the $(i j)$-th element of the inverse
constraint matrix where $\psi^{i}(q,p)$ is a set of Second Class
constraints, then the Dirac bracket between two generic variables
$[A(q,p), B(q,p)]_D$ is given by
\be [A, B]_D = [A, B] - [A,
\psi_{\rho}^{i}] ([\psi_{\rho}^{i}, \psi_{\sigma}^{j}]^{-1})
[\psi_{\sigma}^{j}, B], \label{ddb} \ee
where $[~~,~~]$ denotes Poisson brackets. Let us now study the
constraint structure of our model (\ref{flag}). {\it{Our aim is to
show that the NC algebra (\ref{covcomm}) proposed by us is
realized by the Dirac brackets in this particle model.}}

Following the standard procedure of First-Order formalism, we
consider $x_{\mu}$ and $p_{\mu}$ as two independent variables and
obtain the two sets of constraints $\psi^1_{\mu}$ and
$\psi^2_{\mu}$:
\be \psi^1_{\mu} \equiv \pi^p_{\mu} + \frac{b}{2}
x_{\mu} - \frac{b}{2} \frac{(x p)}{(\alpha p)} \alpha_{\mu} $$$$
\psi^2_{\mu} \equiv \pi^x_{\mu} - \frac{b}{2} \frac{(x p)}{(\alpha
x)} \alpha_{\mu} - (1-\frac{b}{2}) p_{\mu}, \label{cons} \ee
where $\pi^x_{\mu}$ and $\pi^p_{\mu}$ are the momenta conjugate to
$x^{\mu}$ and $p^{\mu}$ respectively, satisfying the commutation
relation
$$[x_{\mu},\pi^x_{\nu}] = [p_{\mu},\pi^p_{\nu}] = g_{\mu \nu}.$$
The constraint matrix and its inverse are given by
\be
[\psi_{\mu}^{i},\psi_{\nu}^{j}] = \left(%
\begin{array}{cc}
  \frac{b}{2 (\alpha p)} (\alpha_{\nu}x_{\mu} - \alpha_{\mu} x_{\nu}) & g_{\mu
\nu}
  + \frac{b}{2(\alpha x)} \alpha_{\nu} x_{\mu}
  - \frac{b}{2(\alpha p)} \alpha_{\mu} p_{\nu} \\
  -g_{\mu \nu} - \frac{b}{2(\alpha x)} \alpha_{\mu} x_{\nu}
  + \frac{b}{2(\alpha p)} \alpha_{\nu} p_{\mu} & \frac{b}{2(\alpha x)}
  (\alpha_{\nu} p_{\mu} - \alpha_{\mu} p_{\nu}) \\
\end{array}%
\right) \label{consmat} \ee

\be [\psi_{\nu}^{i},\psi_{\sigma}^{j}]^{-1} = \left(%
\begin{array}{cc}
  \frac{b}{2(\alpha x)}
  (\alpha_{\sigma} p_{\nu} - \alpha_{\nu} p_{\sigma}) & -g_{\nu \sigma} +
  \frac{b}{2(\alpha x)} \alpha_{\nu} x_{\sigma}
  - \frac{b}{2(\alpha p)} \alpha_{\sigma} p_{\nu} \\
  g_{\nu \sigma}
  - \frac{b}{2(\alpha x)} \alpha_{\sigma} x_{\nu}
  + \frac{b}{2(\alpha p)} \alpha_{\nu} p_{\sigma} &
  \frac{b}{2 (\alpha p)} (\alpha_{\sigma}x_{\nu} - \alpha_{\nu} x_{\sigma}) \\
\end{array}%
\right). \label{consmatinv} \ee
Thus the constraints (\ref{cons}) are Second Class, that is, they
do not commute between themselves under the Poisson bracket. For
this $DISIM_b(2)$ invariant VSR scenario, using (\ref{consmatinv})
we obtain the following Dirac brackets
\be [x_{\mu},p_{\nu}]_D =
\frac{b}{2 (\alpha p)} (\alpha_{\nu} x_{\mu} - \alpha_{\mu}
x_{\nu})~~~,~~~[p_{\mu},p_{\nu}]_D = \frac{b}{2 (\alpha
x)}(\alpha_{\nu} p_{\mu} - \alpha_{\mu} p_{\nu}) $$$$
[x_{\mu},p_{\nu}]_D = g_{\mu \nu} - \frac{b}{2 (\alpha x)}
\alpha_{\nu} x_{\mu} + \frac{b}{2 (\alpha p)} \alpha_{\mu} p_{\nu}
\label{db} \ee
which is exactly of the same form as (\ref{covcomm}). Thus we are
able to provide a first-order Lagrangian (\ref{flag}) whose
kinetic part gives rise to a NC structure between the
phase space variables (\ref{covcomm}). Obviously this
NC structure correctly reproduces the relevant sector
of the $DISIM_b(2)$ algebra given by (\ref{algibbons}), as
discussed in the previous section.

This concludes the first part of our objective as we have been
able to provide a particle model that is a dynamical realization
of the $DISIM_b(2)$ symmetry. The NC phase space structure emerges
inherently and need not be imposed in an {\it{ad hoc}} way. Note
that the dispersion relation in (\ref{flag}), obtained from the
first-order Lagrangian (\ref{flag}), is different
from the one (\ref{g2}) derived in \cite{gibbons}. However, this
new dispersion relation and symplectic structure (\ref{covcomm})
conspire to reproduce the action (\ref{gibblag}) given in
\cite{gibbons}. This is explicitly established in the following
section, where we provide the equivalent coordinate space
second-order (or Nambu-Goto) Lagrangian.

\section{Particle Lagrangian: Second-Order Form}

In the previous section, we have proposed a first-order particle
Lagrangian compatible with the $DISIM_b(2)$ invariant VSR theory.
We proceed further to get a second-order Lagrangian. The
Euler-Lagrange equations of motion for $x_{\mu}$ and $p_{\mu}$
obtained from the Lagrangian $L$ in (\ref{flag}) are respectively
given by:
\be \dot{p}_{\mu} + \frac{b ((\dot{x} p) + (x
\dot{p}))}{2 (\alpha x)} \alpha_{\mu} - \frac{b (\alpha
\dot{x})}{2 (\alpha x)} p_{\mu} - \frac{b (\alpha \dot{p})}{2
(\alpha p)} p_{\mu} + \frac{\lambda b (\alpha p)}{2 (\alpha x)}
p_{\mu} - \frac{\lambda b (x p) (\alpha p)}{2 (\alpha x)^2}
\alpha_{\mu} = 0 \label{delx} \ee
\be \dot{x}_{\mu} + \frac{b (\alpha \dot{x})}{2 (\alpha x)}
x_{\mu} + \frac{b (\alpha \dot{p})}{2 (\alpha p)} x_{\mu} -
\lambda p_{\mu} - \frac{\lambda b (\alpha p)}{2 (\alpha x)}
x_{\mu} - \frac{\lambda b (x p)}{2 (\alpha x)} \alpha_{\mu} -
\frac{b ((\dot{x} p) + (x \dot{p}))}{2 (\alpha p)} \alpha_{\mu}
$$$$ + \frac{\lambda A b (-\alpha p)^{\frac{b-1}{b+1}}}{1+b}
\alpha_{\mu} = 0. \label{delp} \ee
Taking the dot product with $\alpha^{\mu}$ of both the equations
(\ref{delx}) and (\ref{delp}) we obtain the following relations
\be (\alpha \dot{x}) = \lambda (\alpha p) \label{dotxp} \ee \be
(\alpha \dot{p}) = 0. \label{dotp} \ee
We also obtain the following two relations from (\ref{delx}) and
(\ref{delp}):
\be (\dot{x} p) + (x \dot{p}) = -\frac{\lambda
A}{1+b} (-\alpha p)^{\frac{2 b}{1+b}} \label{tot} \ee
\be (x p) = \frac{(x \dot{x})}{\lambda (1+\frac{b}{2})} + \frac{A
b (-\alpha \dot{x})^{\frac{b-1}{b+1}} \lambda^{\frac{1-b}{1+b}}
(\alpha x)}{2 (1+b)(1+\frac{b}{2})}. \label{xp} \ee
Substituting the relations (\ref{dotxp}), (\ref{dotp}),
(\ref{tot}) and (\ref{xp}) in (\ref{delp}), we finally obtain the
expression for the momenta $p_{\mu}$ as
\be p_{\mu} =
\frac{\dot{x}_{\mu}}{\lambda} - \frac{b (x \dot{x})
\alpha_{\mu}}{2 \lambda (1 + \frac{b}{2})(\alpha x)} + \frac{A b
(-\alpha \dot{x})^{\frac{b-1}{b+1}} \lambda^{\frac{1-b}{1+b}}
\alpha_{\mu}}{2(1+\frac{b}{2})(1+b)}. \label{p} \ee
Putting back this expression for $p_{\mu}$ in the Lagrangian $L$
(\ref{flag}) we obtain the second-order Lagrangian in terms of
$x_{\mu}$, $\dot{x}_{\mu}$ and $\lambda$:
\be L =
\frac{\dot{x}^2}{2 \lambda} - \frac{A (-\alpha
\dot{x})^{\frac{2b}{1+b}} \lambda^{\frac{1-b}{1+b}}}{2}.
\label{slag} \ee
Using the equation of motion for $\lambda$ obtained from
(\ref{slag}), we eliminate $\lambda$ from the Lagrangian. The
final expression for the second-order Lagrangian becomes
\be L = -A^{\frac{1+b}{2}} (1-b)^{\frac{b-1}{2}} (1+b)^{-\frac{1+b}{2}}
(-\dot{x}^2)^{\frac{1-b}{2}} (-\alpha \dot{x})^b. \label{slagfa}
\ee
Using the definition of conjugate momenta $$ p_{\mu} =
\frac{\partial L}{\partial \dot{x}^{\mu}},$$ obtained from the
Lagrangian (\ref{slagfa}), the mass-shell condition turns out to
be:
\be p^2 + A (-\alpha p)^{\frac{2 b}{1+b}} = 0.
\label{massshella} \ee
This mass-shell condition (\ref{massshella}) is invariant under
$DISIM_b(2)$ algebra.

If we identify the arbitrary parameter $A$ as $$ A =
m^{\frac{2}{1+b}} (1+b) (1-b)^{\frac{1-b}{1+b}},$$ the Lagrangian
(\ref{slagfa}) takes the form
\be L = -m
(-\dot{x}^2)^{\frac{1-b}{2}} (-\alpha \dot{x})^b, \label{slagf}
\ee
as given in \cite{gibbons}. From the Lagrangian (\ref{slagf}), we
recover the $DISIM_b(2)$ invariant mass-shell condition,
\be p^2 +
m^2(1-b^2)\left(\frac{(-\alpha p)}{m(1-b)}\right)^{\frac{2
b}{1+b}} = 0. \label{massshell} \ee
This is the modified dispersion relation proposed by Gibbons et al
in \cite{gibbons}.

\section{Conclusion}

In this paper, we have considered $DISIM_b(2)$ introduced in
\cite{gibbons}, which is a deformation of a particular VSR group,
$ISIM(2)$, the largest possible subgroup of the Poincar\'e group
compatible with VSR theory. We invoke a non-commutative spacetime
structure in this $DISIM_b(2)$ invariant VSR scenario, where the
momentum algebra becomes non-commutative. We also propose a
first-order Lagrangian that, after performing the Dirac constraint
analysis, reproduces the non-commutative Lie bracket structure
between the phase space variables. Further, we obtained a
second-order Lagrangian which is of the Finsler form suggested in
\cite{bogos, gibbons}.

The present work suggests that it is possible to attribute a
non-commutative momentum algebra with a Finslerian spacetime. But
it should be noted that these momenta are not the translation
generators since $[x_{\mu}, p_{\nu}]$ contains an extension.

\vspace{.5cm}

{\it{\textbf{Acknowledgements:}}} S.G. would like to thank
Professor Gibbons for discussions on VSR during his visit to
DAMTP. S.G. also thanks Department of Mathematics, University of
Cagliari, where a major part of the work was carried out.

\end{document}